\documentclass[preprint,12pt]{emulateapj}

\usepackage{natbib}

\usepackage{graphicx}
\usepackage[space]{grffile}
\usepackage{latexsym}
\usepackage{amsfonts,amsmath,amssymb}
\usepackage{url}
\usepackage[utf8]{inputenc}
\usepackage{fancyref}
\usepackage{hyperref}
\hypersetup{colorlinks=false,pdfborder={0 0 0},}
\usepackage{textcomp}
\usepackage{longtable}
\usepackage{multirow,booktabs}

\newcommand{\kep}{{K2-3}}

\usepackage{xspace}
\usepackage{amsmath}
\newcommand{\RH}{\ensuremath{R_H}}
\newcommand{\Rp}{\ensuremath{R_P}}
\newcommand{\Min}{\ensuremath{M_\mathrm{in}}}
\newcommand{\Mout}{\ensuremath{M_\mathrm{out}}}
\newcommand{\Mstar}{\ensuremath{M_{\star}}}
\newcommand{\ain}{\ensuremath{a_\mathrm{in}}}
\newcommand{\aout}{\ensuremath{a_\mathrm{out}}}
\newcommand{\Mearth}{\ensuremath{M_{\oplus}}}
\newcommand{\Rearth}{\ensuremath{R_{\oplus}}}
\newcommand{\Dbc}{\ensuremath{\Delta_{\rm{bc}}}}
\newcommand{\Dcd}{\ensuremath{\Delta_{\rm{cd}}}}
\newcommand{\Din}{\ensuremath{\Delta_\mathrm{in}}}
\newcommand{\Dout}{\ensuremath{\Delta_\mathrm{out}}}
\newcommand{\Teff}{\ensuremath{T_\mathrm{eff}}}

\shorttitle{A nearby M star with three transiting super-Earths from K2}
\shortauthors{Crossfield et al.}

\begin{document}


\title{\kep: A nearby M star with three transiting super-Earths discovered by K2}


\author{
Ian J. M. Crossfield\altaffilmark{1}$^,$\altaffilmark{12},
Erik Petigura\altaffilmark{2},
Joshua E. Schlieder\altaffilmark{3}$^,$\altaffilmark{13},
Andrew W. Howard\altaffilmark{4},
B.J. Fulton\altaffilmark{4},
Kimberly M. Aller\altaffilmark{4},
David R. Ciardi\altaffilmark{5},
S\'ebastien L\'epine\altaffilmark{6},
Thomas Barclay\altaffilmark{3},
Imke de Pater\altaffilmark{2},
Katherine de Kleer\altaffilmark{2},
Elisa V. Quintana\altaffilmark{3},
Jessie L. Christiansen\altaffilmark{5},
Eddie Schlafly\altaffilmark{7},
Lisa Kaltenegger\altaffilmark{11},
Justin R. Crepp\altaffilmark{8},
Thomas Henning\altaffilmark{7},
Christian Obermeier\altaffilmark{7},
Niall Deacon\altaffilmark{9},
Lauren M. Weiss\altaffilmark{2},
Howard T. Isaacson\altaffilmark{2},
Brad M. S. Hansen\altaffilmark{10},
Michael C. Liu\altaffilmark{4},
Tom Greene\altaffilmark{3},
Steve B. Howell\altaffilmark{3},
Travis Barman\altaffilmark{1},
Christoph Mordasini\altaffilmark{7}
}

\altaffiltext{1}{Lunar \& Planetary Laboratory, University of Arizona Lunar,
1629 E. University Blvd., Tucson, AZ, USA, \href{mailto:ianc@lpl.arizona.edu}{ianc@lpl.arizona.edu}}
\altaffiltext{2}{Astronomy Department, University of California, Berkeley, CA, USA}
\altaffiltext{3}{NASA Ames Research Center, Moffett Field, CA, USA}
\altaffiltext{4}{Institute for Astronomy, University of Hawaii, 2680 Woodlawn Drive, Honolulu, HI, USA}
\altaffiltext{5}{NASA Exoplanet Science Institute, California Institute of Technology,770 S. Wilson Ave., Pasadena, CA, USA}
\altaffiltext{6}{Department of Physics \&\ Astronomy, Georgia State University, Atlanta, GA, USA}
\altaffiltext{7}{Max-Planck Institut f\"ur Astronomie, K\"onigstuhl 17, Heidelberg, Germany}
\altaffiltext{8}{Department of Physics, University of Notre Dame, 225 Nieuwland Science Hall, Notre Dame, IN, USA}
\altaffiltext{9}{University of Hertfordshire, College Lane, AL10 9AB, Hatfield, UK}
\altaffiltext{10}{Department of Physics \& Astronomy, University of California Los Angeles, Los Angeles, CA, USA}
\altaffiltext{11}{Department of Astronomy, Cornell University, 122 Sciences Drive, Ithaca, NY, USA}
\altaffiltext{12}{NASA Sagan Fellow}
\altaffiltext{13}{NASA Postdoctoral Program Fellow}


%\altaffiltext{1}{Visiting Astronomer, Cerro Tololo Inter-American Observatory.
%CTIO is operated by AURA, Inc.\ under contract to the National Science
%\altaffiltext{2}{Society of Fellows, Harvard University.}
%\altaffiltext{3}{present address: Center for Astrophysics,
%    60 Garden Street, Cambridge, MA 02138}
%\altaffiltext{4}{Visiting Programmer, Space Telescope Science Institute}
%\altaffiltext{5}{Patron, Alonso's Bar and Grill}


\begin{abstract}
  Small, cool planets represent the typical end-products of planetary
  formation. Studying the architectures of these systems, measuring
  planet masses and radii, and observing these planets' atmospheres
  during transit directly informs theories of planet assembly,
  migration, and evolution. Here we report the discovery of three
  small planets orbiting a  bright ($K_s = 8.6$~mag) M0 dwarf
  ~using data collected as part of K2, the new ecliptic survey using the
  re-purposed {\emph{Kepler}} spacecraft. Stellar spectroscopy and K2
  photometry indicate that the system hosts three transiting planets
  with radii 1.5 -- 2.1 $\Rearth$, straddling the transition region
  between rocky and increasingly volatile-dominated compositions. With
  orbital periods of 10--45~days the planets receive just
  1.5--10$\times$ the flux incident on Earth, making these some of the
  coolest small planets known orbiting a nearby star; planet d is
  located near the inner edge of the system's habitable zone. The
  bright, low-mass star makes this system an excellent laboratory to
  determine the planets' masses via Doppler spectroscopy and to
  constrain their atmospheric compositions via transit
  spectroscopy. This discovery demonstrates the ability of K2
  and future space-based transit searches to find many fascinating 
  objects of interest.

\end{abstract}

\keywords{\kep --- techniques: photometric ---
techniques:~spectroscopic --- eclipses}

\bibliographystyle{apj}

\section{Introduction}

Surveys for new planets demonstrate that small, low-mass planets are
common around FGK stars 
% and planets occur with increasing frequency toward longer-period orbits 
\citep{howard:2010,howard:2012}.  \cite{petigura:2013b} used {\em
  Kepler} data to measure the frequency of Earth-sized planets in
Earth-like orbits to be 5--20\%. Such small planets with moderate
insolation levels (the stellar energy received by the planet at the
top of any atmosphere) are of considerable interest for their ability
to host Earth-like atmospheres that could potentially support life.

{M dwarfs offer a shortcut to observing rocky and potentially
  habitable planets}.  Compared to nearby Sunlike stars, planets
around M dwarfs are easier to find with transits or radial velocities
(RV), they occur more frequently \citep{howard:2012}, and their
atmospheres are easier to study when transiting
\citep{stevenson:2010,kreidberg:2014}.  Planets transiting M dwarfs
offer the best opportunity to study habitability and constrain models
of rocky planet assembly and migration
\citep[][]{swift:2013,hansen:2014} and of planetary atmospheres
\citep{kaltenegger:2011,rodler:2014}.  Multi-planet M dwarf systems
are even more exciting, both because such candidates are extremely
unlikely to result from astrophysical false positives
\citep{lissauer:2012} and because they allow for studies of
comparative planetology \citep{muirhead:2012a} with identical initial
conditions (i.e., formation in the same natal disk).  However,
relatively few confirmed transiting planets (and fewer multiple
systems) are known around M dwarfs, and the (because {\em Kepler}'s
prime mission targeted just 3900 late-type dwarfs) the prevalence of
planets around M dwarfs is less well constrained than around Sunlike
stars \citep{dressing:2013}.

We are using {K2}, the continuing mission of NASA's {\em Kepler}
spacecraft \citep{howell:2014}, to target thousands of M dwarfs in
each K2 field to find new, small planets orbiting these stars.  K2's
80-day campaigns are ideally suited to finding large numbers of small,
cool planets around M dwarfs, out to semimajor axes in the stars'
habitable zones \citep[e.g.,][]{kopparapu:2014}.  In addition, some of
K2's M-dwarf planets orbit stars bright enough for atmospheric
characterization via JWST transmission or emission spectroscopy
\citep{kaltenegger:2009,batalha:2013,beichman:2015}.

Here, we present the discovery of a new multi-planet system orbiting a
bright M dwarf (\kep, PMI11293-0127, UCAC4 443-054906, PPMX
112920.3-012717).  We describe our analysis of the K2 photometry and
of supplementary imaging and spectroscopic data in Sec.~\ref{sec:obs}.
In Sec.~\ref{sec:conclusions} we present the results of our analysis
of \kep 's properties and discuss the potential for future
observations of this and other systems discovered by K2.

\section{Observations and Analysis}
\label{sec:obs}
We identified the high proper motion star PMI11293-0127 as a target
for our Campaign 1 proposal (GO1036\footnote{The star was also
  identified in programs GO1006, GO1050, GO1052, GO1036, GO1075,
  GO1059, and GO1063.}, PI Crossfield) from the SUPERBLINK proper
motion survey \citep{lepine:2005,lepine:2011}.  We identified the star
as a probable nearby M dwarf based on a color and proper motion
selection scheme and selecting all targets with $(V-J)>2.5$, $V+5 \log
\mu +5 < 10$, and $(6V-7J-3) < 5 \log \mu$, where $\mu$ is the proper
motion.  The star matched
the source \kep\ in the Kepler input catalog \citep{huber:2014}. K2
then observed this target in long-cadence mode during C1, covering
30~May to 21~Aug 2014. Target properties, including optical and NIR photometry from APASS \citep{henden:2012}, 2MASS \citep{skrutskie:2006}, and WISE \citep{wright:2010} are summarized in
Table~\ref{tab:obs}).

\hspace{-1in}
\begin{deluxetable}{l l l }[bt]
\tabletypesize{\scriptsize}
\tablecaption{  Stellar Parameters of \kep \label{tab:obs}}
\tablewidth{0pt}
\tablehead{
\colhead{Parameter} & \colhead{Value} & \colhead{Source}
}

\startdata
\multicolumn{3}{l}{\hspace{1cm}Identifying information} \\
$\alpha$ R.A. (hh:mm:ss) & 11:29:20.388 & \\
$\delta$ Dec. (dd:mm:ss) & -01:27:17.23 & \\
EPIC ID & 201367065 & \\
2MASS ID & 11292037-0127173 & 2MASS  \\
\multicolumn{3}{l}{\hspace{1cm}Photometric Properties} \\
B (mag)..........  & 13.52  $\pm$ 0.06 & APASS  \\
V (mag)..........  & 12.17  $\pm$      0.01 & APASS \\
g (mag) .......... & 12.871 $\pm$ 0.030 & APASS \\
r (mag) .......... & 11.582 $\pm$ 0.020 & APASS \\
i (mag)........... & 10.98 $\pm$ 0.17 & APASS \\
J (mag)..........  & 9.421  $\pm$ 0.027 & 2MASS\\
H (mag) .........  & 8.805  $\pm$  0.044 & 2MASS\\
Ks (mag) ........  & 8.561  $\pm$ 0.023 & 2MASS\\
W1 (mag) ........ & 8.443  $\pm$ 0.022 & AllWISE  \\
W2 (mag) ........ &      8.424  $\pm$ 0.019& AllWISE\\
W3 (mag) ........ & 8.322 $\pm$      0.021 & AllWISE \\
\multicolumn{3}{l}{\hspace{1cm}Spectroscopic and Derived Properties} \\
$\mu_{\alpha}$ (mas~yr$^{-1}$) & 88.3 $\pm$ 2.0 & \cite{zacharias:2012} \\
$\mu{\delta}$ (mas~yr$^{-1}$) & -73.6 $\pm$ 2.7 & \cite{zacharias:2012} \\
Barycentric rv (km~s$^{-1}$) & 32.6 $\pm$ 1 & APF, this paper \\
Distance (pc) & $45 \pm 3$ & this paper\\
EW (H$\alpha$) (\AA) & 0.38 $\pm$ 0.06 & EFOSC, this paper \\
Age (Gyr) & $\gtrsim$1 & EFOSC, this paper \\
Spectral Type & M0.0$\pm$0.5V & This paper.\\
$[$Fe/H$]$ & -0.32 $\pm$ 0.13 & SpeX, this paper\\
\Teff\ (K) & 3896 $\pm$  189 & SpeX, this paper\\
$M_*$ ($M_\odot$) & 0.601 $\pm$ 0.089& SpeX, this paper \\
$R_*$ ($R_\odot$) & 0.561 $\pm$ 0.068& SpeX, this paper \\
\enddata
\end{deluxetable}

\subsection{K2 Photometry}
\subsubsection{Extracting the Photometry}

We extracted the photometry \kep\ from the pixel data, which we downloaded from the MAST. Because K2 only has two functional reaction wheels, the telescope cannot maintain the 50-millipixel pointing precision achieved during the prime mission. The dominant drift is roll around the telescope boresight. When the spacecraft reaches a pre-determined limit the spacecraft corrects this roll with a thruster fire. As the spacecraft rolls, stars move over different pixels having different sensitivities. Thus, motion of the star results in apparent changes in stellar brightness.

Because a target star traces out similar paths during each roll of the spacecraft, it is possible to separate out variations in stellar brightness that are roll angle dependent, and to remove these variations from the photometry. Our extraction pipeline draws heavily on the work of \cite{Vanderburg14}. We begin by computing the median flux for each frame and adopt this value as the background flux level. The background flux is subtracted out on a frame by frame basis. We compute the raw photometry, ${F_\mathrm{SAP}}$, by summing the flux within a soft-edged circular aperture centered around the target star. We compute the row and column centroids within the aperture.

On short timescales, spacecraft roll is the dominant motion term and can be described by a single variable. We identify the roll direction by computing the principle components of the row and column centroids, $x'$ and $y'$. We fit for a function that relates ${F_\mathrm{SAP}} $ to $x'$. We describe this trend by ${F_\mathrm{SAP}}$ = GP($x'$), where GP is a Gaussian process having a correlation matrix given by a squared exponential kernel. Fitting the GP($x'$) is an iterative process where outliers are identified and removed and the hyperparameters associated with the squared exponential kernel are adjusted to yield the minimum residual RMS.

The algorithm described in \cite{Vanderburg14} was developed for the
K2 engineering campaign (C0), where the time baseline was short enough
that drifts in stellar position along the $y'$ direction could be
ignored. During 80~day period of C1 observations, stars moved enough
along the $y'$ direction that the GP($x'$) determined using data early
in the campaign was no-longer an appropriate description of the
position-dependent flux variations. Adopting an approach described in
\cite{Vanderburg14arXiv}, we divided the C1 observations in to six
nearly equal segments and performed the 1D decorrelation approach
described above on each segment individually. The entire procedure
described above is repeated for different aperture radii (2, 3, 4, 5,
6, and 7~pixels). We select the aperture size that minimizes the
calibrated RMS. For \kep, a circular aperture with a 4 pixel radius
yielded the best calibrated photometry (which is available as an
electronic supplement to this paper). Our photometry suggests that
\kep\ may exhibit photometric variations of $\lesssim 1\%$ on
week-to-month timescales, but K2's long-term stability is constrained
sufficiently poorly that we cannot claim evidence for periodic
modulation  indicative stellar rotation.

\subsubsection{Transit Detection}
\begin{figure*}[ht!]
\begin{center}
\includegraphics[width=7in]{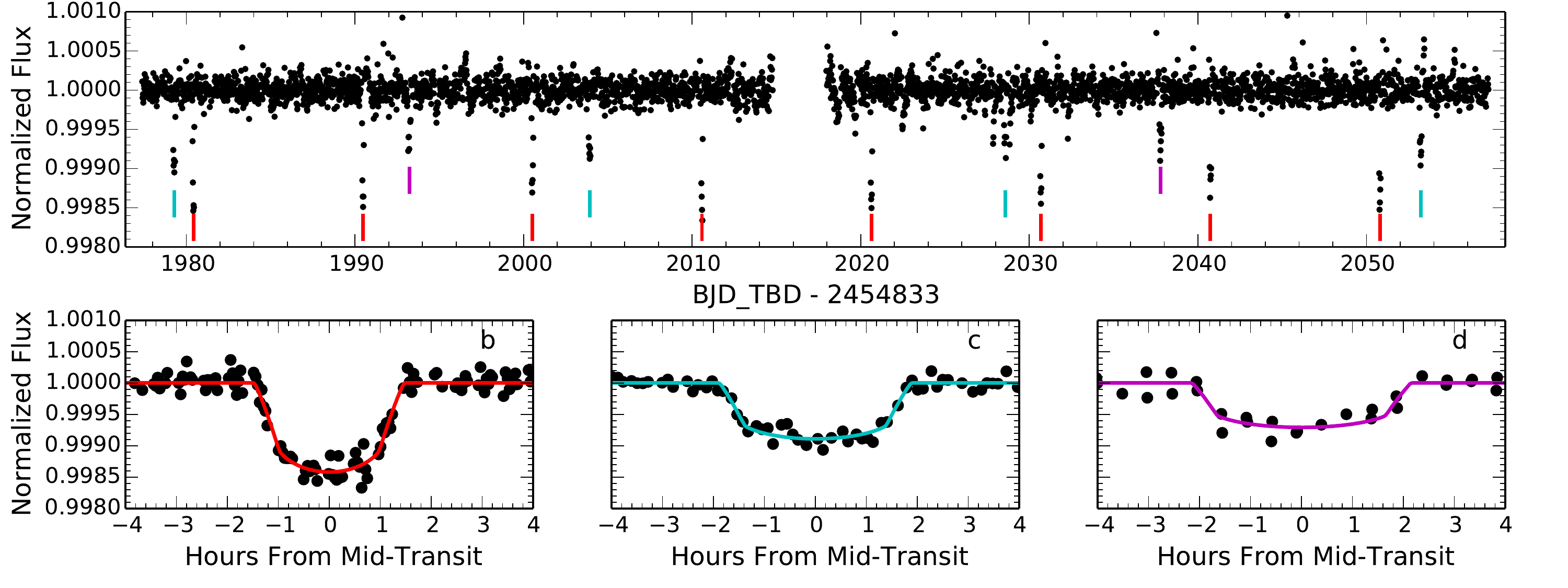}
\caption{\label{fig:fits} {\em Top}: Calibrated K2 photometry for \kep.  Vertical ticks indicate the locations of each planets' transits.  {\em Bottom}: Phase-folded photometry and best-fit light curves for each planet.  }
\end{center}
\end{figure*}

We searched through the calibrated and detrended photometry (shown in
Fig.~\ref{fig:fits}a) using the TERRA algorithm described in
\cite{petigura:2013b}. TERRA identified a transit candidate having
$P$~=~10.056~days and SNR~=~59. We fit this candidate with a
\cite{mandel:2002} model and subtracted the best fit model from the
photometry. We reran TERRA on the photometry with the $P$~=~10.056~day
candidate removed. We found a second candidate having
$P$~=~24.641~days and SNR~=~30. Again we removed the best-fitting
model. TERRA did not find any additional transits, but a $\sim$45-day
candidate was identified by eye (TERRA currently requires 3 detected
transits, and thus was not sensitive to the longest period candidate
which only transits twice during C1).  We fit each of these two
transits individually and find consistent transit parameters,
supporting the hypothesis that they result from a single planet. At
half of this period a third transit would occur in C1's data gap (see
Fig.~\ref{fig:fits}), but this would give the outer two planets a
period ratio of just 1.1. The previous record-holder for a close
period ratio is the Kepler-36 system
\citep[][]{carter:2012,winn:2014}, whose two planets exhibit a
considerably larger period ratio of 1.17 and transit timing variations
of many hours. It is unlikely that such an unusual system would lie
just 45~pc away; in addition, our dynamical analysis (described below)
indicates that this period ratio would be dynamically unstable.  We
therefore conclude that the third planet's period is $\sim$45~d.

\subsection{Target Validation and Stellar Spectroscopy}
We conducted a number of pixel-level diagnostics and observed \kep\
using several spectrographs to constrain the stellar properties. These
observations are described below.  The reduced spectra are attached as
an electronic supplement to this paper, and the derived parameters
are listed in Table~\ref{tab:obs}.

\subsubsection{Pixel-Level and Photometric Data Validation}
Experience over the last decade shows that transit-like signals must
be validated to ensure that they arise from true planets, not
``false-positive'' configurations such as background eclipsing
binaries blended with foreground stars
\citep[e.g.,][]{torres:2004,torres:2011}.  We therefore implement a large
number of tests on the pixel-level data and extracted photometry to
identify and weed out these false positives. 

Once transit-like events are identified, TERRA runs a suite of
diagnostics to distinguish planets from phenomena like eclipsing
binaries, starspots, and other periodic stellar variability. We
subject targets passing this first step to an extensive battery of
further tests which search for blends using an examination of centroid
motions in and out of transit, difference imaging analyses, and
construction of pixel correlation images \citep{bryson:2013}. Though
we are still learning how to optimally tune these tests to account for
K2's few-pixel pointing variations, validation results for large
numbers of targets indicates that the transit-like events identified
with \kep\ occur within roughly one pixel (4'') of the target star.
When combined with our seeing-limited and adaptive optics imaging
described below, as described in Sec.~\ref{sec:fp} we find that
\kep's transits are far more likely to be explained by a multi-planet
system than by nonplanetary phenomena.

\subsubsection{Optical and Infrared Spectroscopy}
We obtained $R\sim1500$ spectra from 0.6--1.0$\mu m$ of \kep\ and a number of calibration objects using NTT/EFOSC2 \citep{buzzoni:1984} on UT 11 Jan 2015 as part of 70-night K2 followup program (PID 194.C-0443, PI Crossfield).  We draw our calibrators from several recent works \citep{boyajian:2012b,pecaut:2013,mann:2013b}.  A forthcoming paper will discuss these efforts; in brief, we bias-subtract and flat-field the data frames, extract spectra using IRAF, and wavelength-calibrate using EFOSC2's internal HeAr lamps.  We achieve a S/N per resolution element of $\sim$100 for \kep\ and somewhat higher for our reference sample.  We  flux-calibrate the extracted spectrum using observations of spectrophotometric standards.

We observed \kep\ on 2015 January 11 UT using the recently refurbished SpeX spectrograph \citep{rayner:2003} on the 3.0m NASA Infrared Telescope Facility (IRTF). The data were taken under clear skies with an average seeing of $\sim$0\farcs7. We observed with the instrument in short cross dispersed mode (SXD) using the 0.3 X 15" slit. This setup provides simultaneous wavelength coverage from 0.7 to 2.5 $\mu$m at a resolution of R$\approx$2000. The extended blue wavelength coverage is a result of the recent chip upgrade SpeX received in July 2014. The target was placed at two positions along the slit and observed in an ABBA pattern for subsequent sky subtraction. The observing sequence consisted of 8 $\times$ 40\,s exposures for a total integration time of 320s. Once the exposures were stacked, this integration time led to a signal-to-noise of $>140$ per resolution element. We obtained standard SpeX calibration frames consisting of flats and arclamp exposures immediately before observing \kep.

\begin{figure*}[ht!]
\begin{center}
\includegraphics[width=7in]{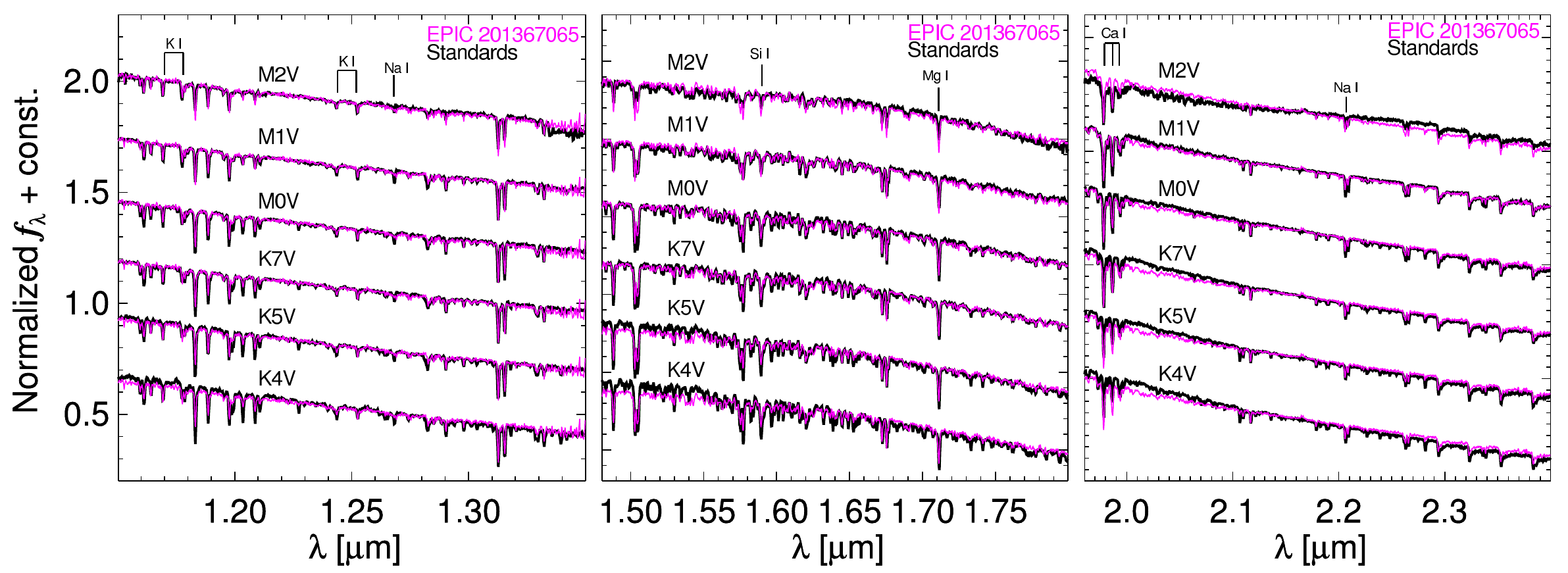}
\caption{\label{fig:spec} Calibrated IRTF/SpeX spectra of our target compared to spectral standards. Stellar parameters are tabulated in Table~\ref{tab:obs}.}
\end{center}
\end{figure*}

The SpeX spectrum was reduced using the SpeXTool software package \citep{cushing:2004}. SpeXTool performs flat-field correction and wavelength calibration from the calibration frames followed by sky subtraction and extraction of the one-dimensional spectrum. Individual exposures of the target were combined using the \textit{xcombspec} routine within SpeXTool. We corrected for atmospheric absorption and performed flux calibration using the A0V-type star HD 97585 which was observed within 20 minutes and 0.015 airmass of the target. A telluric correction spectrum was constructed from the spectrum of the A0V using the \textit{xtellcor} package \citep{vacca:2003} and applied to the spectrum of \kep. This package also performs flux calibration. Separate, telluric-corrected SpeX orders were combined and flux matched into a continuous spectrum using the \textit{xmergeorders} routine. To minimize errors in the spectral slope due to changes in seeing, guiding, and differential refraction, we aligned the slit with the parallactic angle and minimized the time between observations of the target and standard star. Prior to performing any spectroscopic analyses, we also applied corrections for the barycentric velocity of the observatory and the measured radial velocity.  The final, calibrated spectrum is shown in Fig.~\ref{fig:spec}.

\subsubsection{Stellar Parameters}

\cite{mann:2013a} motivate a set of temperature sensitive spectral
indices spanning the visible, J-, H-, and K-bands that are calibrated
using the M dwarf sample of \cite{boyajian:2012b} with
interferometrically measured radii. We used these indices to estimate
the temperature of \kep. We calculate the mean of the temperatures
from each of the three NIR band indices and their rms scatter and find
\Teff\ = 3896$\pm$117 K ($\pm$148 K systematic error, $\pm$189 K total
error). This range of effective temperatures is consistent with that
main-sequence M0 dwarfs of spectral type K8V to M0V
\citep{pecaut:2013}, using the modified system which incorporates
subtypes K8V and K9V between K7V and M0V.

% Many authors have recently proposed methods to accurately ($\sim$0.1 dex) determine M dwarf metallicities from NIR spectra, which mainly focus on the EW's of metallcity sensitive lines in the NIR, particularly Na I, Ca I, K I alkali lines in the H- and K-bands. 
We adopt the metallicity calibration of \cite{mann:2013b} to remain consistent with our methods for determining \Teff, and other parameters. We use custom IDL software provided by A. Mann\footnote{\url{https://github.com/awmann/metal}} to calculate the metallicity in in the visible, J-, H-, and K-bands following the calibrations of \cite{mann:2013b}. Since our SpeX spectrum does not extend $<0.7$ $\mu$m, we do not use the visible band calibrations. Following the suggestion of \cite{mann:2013b}, we also discard the J-band metallicity, which is often an outlier. Our final metallicity is the mean of those measured from the H- and K-band relations and the error is the quadrature sum of the measurement error and systematic error in each band. We find [Fe/H] = -0.32$\pm$0.13. Thus, \kep's metallicity is sub-solar, broadly consistent with many other nearby, field-age, M dwarfs.

\cite{mann:2013a} provide empirical calibrations to calculate the radii, masses, and luminosities given the \Teff\ of an M dwarf. We estimate these additional fundamental parameters again using IDL software written by A.~Mann\footnote{\url{https://github.com/awmann/Teff_rad_mass_lum}} to calculate radius, mass, and luminosity and their associated errors using the relations detailed in \cite{mann:2013b}. Using the most conservative \Teff\ errors, we calculate $R_*$ = 0.561$\pm$0.068 $R_\odot$ and $M_*$ = 0.601$\pm$0.089 $M_\odot$.
%, and $L_*$ = 0.063$\pm$0.027 $L_{sun}$. 
These values, and the other fundamental parameters of the star, are tabulated in Table \ref{tab:obs} and are used for subsequent estimates of the individual planet properties.

Independent of these parameters, we also assign a spectral type to
this star using molecular band heads in our optical and NIR
spectra. In the optical, the TiO5, CaH2, and CaH3 indices
\citep{reid:1995,gizis:1997} are calibrated for the earliest M dwarfs
\citep{lepine:2003} and avoid regions of the spectrum with heavy
telluric contamination. Following the most recent spectral type
calibrations of these indices by \cite{lepine:2013}, our EFOSC
spectrum yields a spectral type of K7.5$\pm$0.5, determined to a
half-subtype scale and assuming a sequence K5-K7-M0 \citep[i.e.,
without the K8 and K9 subdivisions of ][]{pecaut:2013}.  In the NIR,
the H$_2$O-K2 index measures water opacity in the K-band, and was
calibrated to a spectral subtype by \cite{rojas-ayala:2012}. We
calculate this index from our SpeX spectrum and estimate a spectral
type of M0.5$\pm$0.5.  \cite{lepine:2013} also provide a calibration
of the $V-J$ color to spectral subtype.  Our target has $V-J=2.75$,
consistent with subtype K7.5 on the scale of \cite{pecaut:2013}.

The spectroscopic and photometric classifications are all consistent,
although the NIR classification is marginally later. Here we average
the optical and infrared results and adopt a spectral type of
M0.0$\pm$0.5V. Using the $riJHK$ photometric calibrations of
\cite{kraus:2007}, we estimate a distance to \kep\ of $45 \pm 3$~pc.

We obtained high-resolution (2'' slit width with the B
decker) spectra of \kep\ with the Levy Spectrometer
\citep{radovan:2010} on the Automated Planet Finder (APF) telescope
\citep{vogt:2014}.  The spectra were reduced using standard
procedures, as described in \citep{fulton:2015}. Inspection of the gravity-sensitive lines confirms that
\kep\ is a high gravity target, consistent with the medium resolution
spectra described above. We do not see any evidence of a second set of
spectral lines, ruling out companions $\sim$2.5~mag fainter than \kep\ at
visible wavelengths.

\subsubsection{Activity, Age, and Membership}
Lines in the Balmer series are associated with magnetic activity in
late-type stars. The strongest line in the series, H$\alpha$ at
6563~\AA, is classically used to asses the activity of M dwarfs and as
a crude indicator of age \citep{west:2004,west:2008}. We therefore
measure the H$\alpha$ equivalent width (EW) as defined by
\cite{west:2011} and \cite{lepine:2013} and find consistent results
using both approaches. We use two different integration regions to
calculate this EW and apply Monte Carlo methods to estimate the
uncertainty in the EW measurements.  We find that EW $= 0.38\pm
0.06\AA$, indicating that \kep\ is a relatively inactive star. We
further investigate possible chromospheric activity in \kep\ by
analyzing its UV emission measured by GALEX \citep{martin:2005}. The
star is a weak near-UV (NUV) emitter and is not detected in the far-UV
(FUV). Its low NUV flux and non-detection in the FUV is consistent
with quiescent emission, similar to other nearby field M dwarfs
\citep{shkolnik:2011}.  The H$\alpha$ absorption, UV fluxes, and lack
of chromospheric activity in an M0 dwarf all indicate an old, field-age
star and translates to a lower age limit of $\sim$1 Gyr
\citep{west:2008}.

We further examined the possibility that \kep\ is young by comparing
its space position ($XYZ$) and kinematics ($UVW$) with those of known
young moving groups. Its 6-dimensional $UVWXYZ$ position is
inconsistent with the well-known nearby young ($\approx$10--100~Myr)
groups summarized by \cite{gagne:2014}, as well as other sparser or
slightly older groups \citep{shkolnik:2009,zuckerman:2013}. To provide
a quantitative estimate, we used the BANYAN~II web tool
\citep{malo:2013,gagne:2014}. BANYAN~II calculates the probability of
an object being a member of a nearby young moving group using Bayesian
inference and the observed proper motion, sky coordinates, radial
velocity and distance. The probability of \kep\ being a
member of one of the known nearby young moving groups is $<$0.1\% given
the sky coordinates, proper motion, and radial velocity. Inclusion of
the photometric distance estimate (and conservatively assuming a 20\%
distance uncertainty) does not change the BANYAN~II results. Thus we
conclude that \kep\ is unlikely to be a member of any of
these young moving groups.

\subsection{Archival and Adaptive Optics Imaging}
\label{sec:ao}
\begin{figure*}[bt!]
\begin{center}
\includegraphics[width=7in]{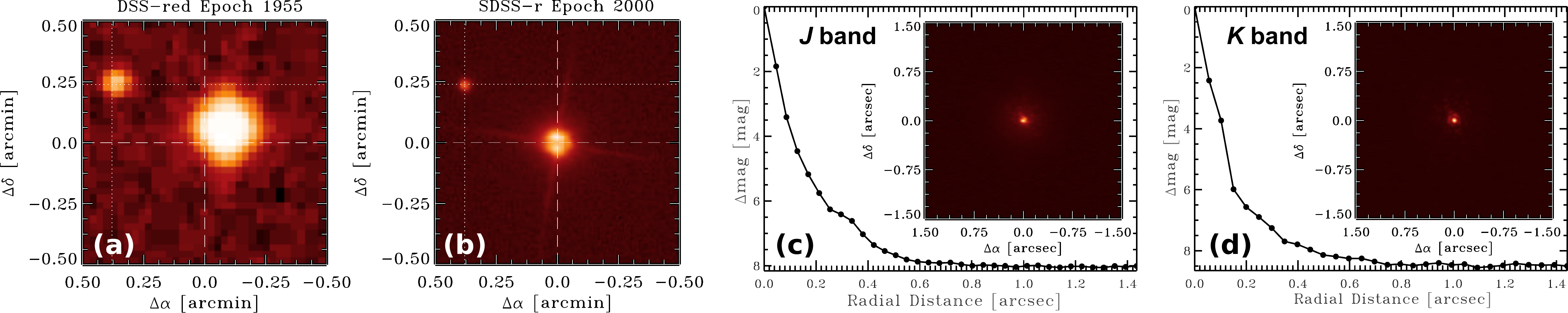}
\caption{\label{fig:keck} We detect no objects within 25'' of \kep:
  in DSS (a); in SDSS (b); and with Keck/NIRC2 in $J$-band (c) and
  $K$-band (d).  }
\end{center}
\end{figure*}

To rule out the presence of a background star being the source of or
diluting the transit events, we compare two epochs of imaging data
from the Digitized Sky Survey (DSS) and the Sloan Digital Sky Survey
\citep[SDSS; ][]{ahn:2012} separated by 45 years.  The data shown in
Fig.~\ref{fig:keck} are the DSS-Red plates with a pixel scale of
1.7\arcsec/pixel taken on 19 April 1955 and the SDSS $r$-band image
with a pixel scale of 0.396\arcsec/pixel taken on 03 March 2000.  The
images are 1~arcminute on a side and clearly show the proper motion of
the primary target.  The nearby star located 27\arcsec\ to the NE is
consistent with zero motion within our astrometric uncertainties; this
star lies outside the photometric aperture applied to the K2
photometry. The primary target, in contrast, displays a clear proper
motion of 6.2\arcsec\ over 45 years, in reasonable agreement with the
measured proper motion \citep{lepine:2011,zacharias:2012}.  In the DSS
image there is no evidence of a background star, and we estimate if a
star is located at the position of the primary target in the Kepler
data, that star must be at least 6 magnitudes (or more) fainter than
the target star.

Near-infrared adaptive optics imaging of \kep\ was obtained at Keck
Observatory on the nights of 2015 January 12 UT and 2015 Janary 16 UT.
Observations were obtained with the $1024\times1024$ NIRC2 array and
the natural guide star system; the target star was bright enough to be
used as the guide star. The data were acquired in the narrow-band
K-band and J-band continuum filters (Kcont and Jcont) using the narrow
camera field of view with a pixel scale of 9.942~mas~pix$^{-1}$; the
atmosphere was less stable on night 1 and only Kcont was acquired on
that night.  A 3-point dither pattern was utilized to avoid the
noisier lower left quadrant of the NIRC2 array.  For both nights, the
3-point dither pattern was observed with 10 coadds and a 1.5
integration time, but on night 1 only 4 frames were acquired for a
total of 60 seconds of on-source exposure time. For night 2, three
full dither patterns were acquired for a total on-source exposure time
of 135 seconds in both Kcont and Jcont filters.  The data from each
night were flatfielded and sky subtracted and the dither positions
were shifted and coadded into a single final image. The final images
from night 2 are shown in Fig.~\ref{fig:keck}c and~d.

For night 1, the target star was measured with a resolution of
$0.07^{\prime\prime}$ (FWHM), but the atmosphere was much more stable
during night 2 and these images have a resolution of
$0.05^{\prime\prime}$ in the Kcont filter and $0.04^{\prime\prime}$ in
the Jcont filter.  No other stars were detected within the
$10^{\prime\prime}$ field of view of the camera; speckles seen in the
Kcont images are not co-spatial with the speckles seen in the Jcont
image, indicating that the speckles are not faint companions.  The
night 2 data were much more sensitive than the night 1 data and we
report the analysis of those data in this work.  In the Kcont filter,
the data are sensitive to stars that have K-band brightness of $\Delta
mag = 2.4$ mag at a separation of $0.05^{\prime\prime}$ and $\Delta
mag = 8.0$ mag at a separation of $0.5^{\prime\prime}$ from the
central star; in the Jcont filter, the data are sensitive to stars
that have J-band brightness of $\Delta mag = 2.0$ mag at a separation
of $0.05^{\prime\prime}$ and $\Delta mag = 7.5$ mag at a separation of
$0.5^{\prime\prime}$ from the central star (see Fig.~\ref{fig:keck}c
and~d).  We estimate the sensitivities by injecting fake sources
with a signal-to-noise of 5 into the final combined images at
distances of N*FWHM from the central source.  The 5$\sigma$
sensitivities, as a function of radius from the star are shown in
Fig.~\ref{fig:keck}c and~d.

\subsection{Light Curve Fitting}
We analyze the photometry using standard Python-based minimizers, the
\texttt{emcee} Markov Chain Monte-Carlo (MCMC) package
\citep{foreman-mackey:2012}, and the JKTEBOP lightcurve code
\citep{southworth:2004,southworth:2011} using numerical integration to
account for our $\sim$30-min cadence.  We fit each planet's transit
separately, after first masking out data taken during the other
planets' transits.

We use the best-fit TERRA parameters to initialize the fits. We assumed a linear limb-darkening relation for the star. Because the data are insufficient to break all degeneracies between the light curve parameters \citep{muirhead:2012a}, we impose Gaussian priors in our analysis. For the limb-darkening parameter $u$, we assume a distribution with center 0.560 and dispersion 0.044; these values correspond to the mean and standard deviation, respectively, of all linear limb-darkening terms tabulated by \citep{claret:2012} that satisfy $3300\le T_{eff} \le3700$~K and $\log_{10} g \ge 4.5$. Using  the spectroscopic parameters presented below (Table~\ref{tab:obs}), we also impose a prior on the stellar density to constrain $R_*/a$ \citep{seager:2003}.  This last point assumes that the planets' orbits are circular, an assumption that future RV measurements will test.

We seed our 60 MCMC chains with values near the best-fit parameters. We assign our data points equal weights, such that the best-fit likelihood equals $-\chi^2/2$. After burn-in we run the MCMC sampler: after each set of 2000 steps, we optimize the fits given by each chain's parameters to check for better fits to the data.  We re-initialize the sampler and re-scale the data weights if we find an improved fit, repeating until all parameters' chains are well-mixed \citep[as indicated by Gelman-Rubin metrics $\le 1.03$;][]{gelman:1992}.   As our final confidence intervals, we use the 15.87\% and 84.13\%
percentiles of each parameters' posterior distribution.  The final distributions are unimodal.  Fig.~\ref{fig:fits} shows the resulting photometry and best-fit models, and Table~\ref{tab:planet} summarizes the final values and uncertainties.

\subsection{Ruling Out False Positives}
\label{sec:fp}
Almost all candidates in {\em Kepler}'s multi-planet systems are {\em
  bona fide} planets \citep{lissauer:2011}, but one pernicious source
of confusion is the possibility of mistaking blended stars each
hosting their own planets for a single multi-planet system.  We
therefore investigated the possibility that \kep\ might be a blend of
multiple stars.  First, we note that \kep 's proper motion (listed in
Table~\ref{tab:obs}) is large enough that optical DSS survey images
reveal no objects at the star's current location (see
Fig.~\ref{fig:keck}a and~b).  Our Keck/NIRC2 images also show no companions
at separations down to a fraction of an arc second (see
Fig.~\ref{fig:keck}c and~d), and so our data validation tests indicate
that the transits must occur around \kep\ and not around some other
nearby star.  Blends involving background eclipsing binaries are thus
strongly disfavored.

The most likely remaining false positive configuration involves a
heirarchical triple system, with a later-type M dwarf close to \kep\
and with its own transiting planet(s) --- but this too is extremely
unlikely.  An M4 dwarf would have $\Delta K_p\approx 2.7$ and so might
be missed in our APF and EFOSC spectra, but the M4 would have $\Delta
K_s \approx 2.0$ \citep{kraus:2007} and so to avoid detection in our
Keck/NIRC2 image it would need to lie at $a\lesssim 2.3$~AU --- while
still needing to host its own $2 \Rearth$ transiting planet.  The
likelihood that \kep\ has such a low-mass stellar companion is
$\sim$0.4 and that such a companion would lie at a projected
separation $<2.3$~AU is $\sim$0.5 \citep{duchene:2013}.  For planet b,
the likelihood of an M dwarf hosting such a planet is $\lesssim0.15$
\citep{dressing:2013}; and the likelihood of it transiting is
$\sim$0.02. Then the likelihood of such a contrived configuration is
just $\sim6\times 10^{-4}$ (1 in 1700), so we eliminate this scenario
as well -- in any case, such a binary would be quickly revealed by
even crude radial velocity measurements.  We therefore conclude that
\kep\ indeed hosts a three-planet system.

\begin{deluxetable*}{l l l l l l}[bt]
\tabletypesize{\scriptsize}
\tablecaption{  Planet Parameters \label{tab:planet}}
\tablewidth{0pt}
\tablehead{
\colhead{Parameter} & \colhead{Units} & \colhead{b} & \colhead{c} & \colhead{d} 
}
\startdata
  $T_{0}$ & $BJD_{TDB} - 2454833$ & $1980.4189^{+0.0011}_{-0.0011}$ & $1979.2786^{+0.0026}_{-0.0027}$ & $1993.2232^{+0.0037}_{-0.0043}$\\
       $P$ &          d & $10.05403^{+0.00026}_{-0.00025}$ & $24.6454^{+0.0013}_{-0.0013}$ & $44.5631^{+0.0063}_{-0.0055}$\\
       $i$ &        deg & $89.28^{+0.46}_{-0.60}$ & $89.55^{+0.29}_{-0.44}$ & $89.68^{+0.21}_{-0.26}$\\
 $R_P/R_*$ &         \% & $3.483^{+0.123}_{-0.070}$ & $2.786^{+0.143}_{-0.083}$ & $2.48^{+0.14}_{-0.10}$\\
  $T_{14}$ &         hr & $2.553^{+0.047}_{-0.044}$ & $3.428^{+0.106}_{-0.097}$ & $3.98^{+0.17}_{-0.15}$\\
   $R_*/a$ &         -- & $0.0343^{+0.0049}_{-0.0020}$ & $0.0193^{+0.0041}_{-0.0014}$ & $0.0127^{+0.0025}_{-0.0010}$\\
       $b$ &         -- & $0.37^{+0.22}_{-0.23}$ & $0.41^{+0.26}_{-0.25}$ & $0.45^{+0.23}_{-0.28}$\\
       $u$ &         -- & $0.560^{+0.041}_{-0.042}$ & $0.557^{+0.043}_{-0.044}$ & $0.563^{+0.041}_{-0.042}$\\
%  $\rho_*$ & g~cm$^{-3}$ & $4646.045^{+899.034}_{-1536.085}$ & $4329.8920^{+1125.0728}_{-1890.2161}$ & $4692.8369^{+1371.3013}_{-1945.1281}$\\
       $a$ &         AU & $0.0769^{+0.0036}_{-0.0040}$ & $0.1399^{+0.0066}_{-0.0073}$ & $0.2076^{+0.0098}_{-0.0108}$\\
     $R_P$ &      $\Rearth$ & $2.14^{+0.27}_{-0.26}$ & $1.72^{+0.23}_{-0.22}$ & $1.52^{+0.21}_{-0.20}$\\
 $S_\mathrm{inc}$ &      $S_\oplus$ & $11.0^{+4.1}_{-3.1}$ & $3.32^{+1.25}_{-0.95}$ & $1.51^{+0.57}_{-0.43}$\\
$T_\mathrm{eq}$ & (K) & $\sim$800 & $\sim$450 & $\sim$300 

\enddata
\end{deluxetable*}

\subsection{System Stability}
Here we investigate the dynamical stability of the three-planet \kep\
system. The planet masses are unconstrained by transit photometry, so we adopt the following mass-radius relationship:
\begin{itemize}
\item $M = \frac{4\pi}{3} R^3 \rho$, where $ \rho = (2.43 + 3.39*(\Rp / \Rearth ))$ g cm$^{-3}$ for $\Rp < 1.5 \Rearth$ \citep{weiss:2014}

\item $M = 2.69 \Mearth \left(\frac{\Rp}{\Rearth} \right)^{0.93} $ \citep{weiss:2014} for $1.5 \Rearth < \Rp < 4.0 \Rearth$

\item $M  = \Mearth \left(\frac{\Rp}{\Rearth}\right)^{2.06} $ for $\Rp > 4.0 \Rearth$ \citep{lissauer:2012}
\end{itemize}
Adopting the above mass-radius relationship we derive masses of 5.3,
4.3, and 4.4~\Mearth\ for planets b, c, and d respectively. We
integrate the system forward in time with the Mercury integration
package \citep{chambers:1999} utilizing the hybrid integrator and
found the system to be stable for the full $2\times10^5$~yr
simulation.

We also evaluate analytically the system's stability. The relevant
length scale for dynamical interactions between planets is the mutual
Hill radius:
\begin{equation}
\RH = \left[\frac{\Min + \Mout}{3 \Mstar} \right]^{1/3} \frac{\ain + \aout}{2}
\end{equation}

where $M$ and $a$ denote mass and semi-major axis, respectively. The subscripts ``in'' and ``out'' correspond to the inner an outer planets respectively. Following
\cite{Fabrycky12}, for each pair of planets, we compute $\Delta =
\left(\aout - \ain \right) / \RH$, the separation between the planets
measured in units of their mutual Hill radii. If two planets begin on
circular orbits, they are stable indefinitely if $\Delta > 2\sqrt{3}
\approx 3.5$ \cite{Gladman93}. In the case of \kep, $\Delta_{\rm{bc}} = 15.9 $ and
$\Delta_{\rm{cd}} = 11.0$. Thus, the two pairs of adjacent planets do
not violate the criterion of Hill stability.

There is no analytic stability criterion for systems having three or more planets \cite{Fabrycky12}. \citet{Fabrycky12} introduce \Din + \Dout, as a heuristic metric for assessing the stability of three planets in triple or higher multiplicity systems. They adopt $\Din + \Dout > 18$ as a heuristic criterion for the stability of three planets, motivated by suites of direct numerical integrations (e.g. \cite{Smith09}). This criterion is empirically supported by the ensemble of systems with three or more transiting planets from the Kepler mission. Among the 413 such systems in \citet{Fabrycky12}, only six had $\Din + \Dout < 18$. For \kep, $\Dbc + \Dcd = 26.9$, and thus has a similar architecture to the ensemble of triple and higher systems discovered during the prime Kepler mission.

\section{Discussion}
\label{sec:conclusions}
Our analysis indicates three small planets orbiting this bright, nearby M dwarf. The planets range in size from $2 \Rearth$ to $1.5 \Rearth$, indicating that they may span the gap between rock-dominated ``Earths''/``super-Earths'' and low-density ``sub-Neptunes'' with considerable volatile content \citep{marcy:2014,rogers:2015,dressing:2015a}.  

The planets' radii imply  masses of roughly 4--5 $M_E$ and Doppler amplitudes of 1.2--2.3 m\,s$^{-1}$, within reach of modern RV spectrographs.  These mass estimates assume that the planets fall on the mean mass-radius relationship, characterized by high densities and rocky compositions for planets smaller than $\sim$1.6 $R_E$.  However, most of the planets with measured masses and $R_p < 1.6 R_E$ have high incident fluxes \citep[e.g.,][]{batalha:2011,howard:2013,pepe:2013}.  The mass-radius relationship is poorly constrained for cool planets that are less likely to be sculpted by thermal evolution and photo-evaporation \citep{lopez:2012}.  Characterizing the mass-radius relationship for these cool, small planets is an important step to learning whether Earth-size planets in the habitable zone also have Earth-like atmospheres.

The planets' receive insolation levels ($S_\mathrm{inc}$) roughly 11, 3.3,
and 1.5$\times$ that of the Earth for planets b, c, and d,
respectively.  Planet d is located at the inner edge of the system's
habitable zone, with $S_\mathrm{inc}=1.51^{+0.57}_{-0.47} S_\oplus$ -- close
to the limits of the empirical habitable zone
\citep[e.g.,][]{kopparapu:2014}-- making this planet a very
interesting potential super-Venus or super-Earth. Because this system
is so close the atmosphere of this planet can be explored in the near
future; depending on atmospheric, cloud, and surface properties liquid water
could potentially persist on planet~c \citep[][but see
\citeauthor{kasting:2014} 2014]{zsom:2013}.

The \kep\ system is a convenient system to measure the atmospheric properties of small, cool planets.  Indeed, the star is a full magnitude brighter than Kepler-138 \citep{kipping:2014}, the previous best system for characterizing cool, nearly Earth-size planets. For cloud-free, hydrogen-dominated atmospheres, we estimate that these planets will show spectral features with amplitudes of $10 H R_p/ R_\star^2$ on the order of 100--200 ppm \citep{miller-ricci:2009}, where $H$ is the atmospheric scale height.  These features would be detectable with current instrumentation on the Hubble Space Telescope \citep{kreidberg:2014}.  Transit features in a heavy atmosphere (e.g., N$_2$, CO$_2$) would be an order of magnitude smaller, and  secondary eclipses will have depths on the order of $(R_p/R_{\star})^2 T_\mathrm{eq} / T_{\star} \sim$  50--150 ppm -- either of these scenarios should be detectable with JWST.  
%For a given cloud-free atmospheric composition, all three planets should present roughly equal-amplitude features when observed with transit spectroscopy \citep{miller-ricci:2009}, and secondary eclipse observations should be feasible with JWST.   
% the estimated S/N of features seen in transmission by JWST will be comparable to that for GJ~1214b with HST.  and 
By allowing us to measure masses and atmospheric conditions for 3 small planets in a single system, \kep\ represents an exciting opportunity to test theories of planet formation and evolution in a single extrasolar laboratory.  

That K2 should reveal such a system in its first full  campaign
demonstrates that the mission will extend {\em Kepler}'s compelling scientific legacy for years to come.  Along with HIP~116454 \citep{vanderburg:2015hip}, the discovery of \kep\ shows that K2 is already finding fascinating new targets for observation with JWST and heralds an era of further unprecedented discoveries in the {\em TESS} era.

\acknowledgements{Acknowledgements} We thank Geoff Marcy, Evan
Sinukoff, and Charles Beichman for helpful conversations, Vishnu Reddy
for swapping SpeX time, and Steve Bryson and our referee
Don Pollacco for useful comments that improved the quality of this
manuscript.  A.\,W.\,H.\ acknowledges NASA grant NNX12AJ23G, and
S.L. acknowledges NSF grant AST 09-08419.  This work made use of the
SIMBAD database (operated at CDS, Strasbourg, France), NASA's
Astrophysics Data System Bibliographic Services, the Authorea
collaborative writing website, the NASA Exoplanet Archive and Infrared
Science Archive, and data products from the Two Micron All Sky Survey
(2MASS), the APASS database, the SDSS-III project, the Digitized Sky
Survey, and the Wide-Field Infrared Survey Explorer.  Portions of this
work were performed at the California Institute of Technology under
contract with the National Aeronautics and Space Administration.  Some
of the data presented herein were obtained at the W.M. Keck
Observatory (which is operated as a scientific partnership among
Caltech, UC, and NASA) and at the Infrared Telescope Facility (IRTF,
operated by UH under NASA contract NNH14CK55B). The authors wish to
recognize and acknowledge the very significant cultural role and
reverence that the summit of Mauna Kea has always had within the
indigenous Hawaiian community.  We are most fortunate to have the
opportunity to conduct observations from this mountain.

{\it Facility:} \facility{Kepler}, \facility{K2}, \facility{IRTF (SPEX)}, \facility{NTT (EFOSC2)}, \facility{APF (Levy)}, \facility{Keck-II (NIRC2)}

\bibliography{converted_to_latex.bib}

\end{document}